\documentclass[options]{JHEP3}
\title{SU(N) fundamental and adjoint confining strings}
\author{Dmitri Antonov\thanks{Permanent address: ITEP, B. Cheremushkinskaya 25, RU-117 218, Moscow, Russia.}\\  
INFN-Sezione di Pisa, Universit\'a degli studi di Pisa,
Dipartimento di Fisica ``E. Fermi'', Via Buonarroti, 2 - Ed. B-C - 56127 Pisa, Italy\\ 
E-mail: \email{antonov@df.unipi.it}}
\abstract{String representations of the Wilson loop are constructed in the 
$SU(N)$-version of compact QED in three and four dimensions. This is done exactly in the case of the fundamental Wilson loop and 
in the large-$N$ limit in the case of the adjoint Wilson loop. Using for concreteness the three-dimensional fundamental case,
it is demonstrated how the resulting $SU(N)$-generalization 
of the so-called theory of confining strings can be 
obtained in various ways. In its weak-field limit (corresponding to the limit of low 
monopole densities), this theory enables one to fix the value of the string tension, 
which cannot be fixed when deduced from the mean magnetic field inside a flat contour (the derivation of this field is also presented).
Moreover, the obtained theory enables one to find also the coupling constants of terms
in the expansion of the nonlocal string effective action, which are higher in the derivatives
than the Nambu-Goto term (some of these terms vanish at the flat surface). In the four-dimensional case with the $\theta$-term,
the critical values of the $\theta$-parameter, at which the problem of crumpling of large world sheets might be solved, are found
in both the fundamental case and in the large-$N$ limit of the adjoint case.
These values are only accessible provided the electric coupling constant is larger than a certain value, in accordance 
with the known fact that confinement in the four-dimensional compact QED holds in the strong-coupling regime.} 
\preprint{IFUP-TH 2003/17}
\dedicated{}
\begin{document}

\section{Introduction}
One of the most powerful methods, aimed at the description of confinement in QCD, is the method 
of Abelian projections~\cite{th} (see ref.~\cite{suN} for the latest developments and ref.~\cite{digiacomo} for reviews). Among various ways to
model the grand canonical ensemble of Abelian-projected monopoles, the most natural one is 
to treat them as the Coulomb plasma, analogous to the one of compact QED~\cite{polpl} (see ref.~\cite{dg} for a review). 
The string description
of the latter in terms of the Kalb-Ramond field, the so-called theory of confining strings, has been proposed
for the 3D-case in ref.~\cite{pol} and then generalized to the 4D-case in ref.~\cite{cristina}.
It should however been mentioned that the ideas to describe confinement in terms of the 
Kalb-Ramond field were put forward almost 30 years ago~\cite{nr}. These ideas have been elaborated 
further on in ref.~\cite{orland}, where the connection between magnetic condensation and the Kalb-Ramond formulation 
of confinement has been pointed out.
Moreover, in ref.~\cite{orland} a non-Abelian generalization of the Kalb-Ramond field
has been proposed and used in an attempt to
construct a string representation of the Wilson loop on the lattice.

A problem remained since the time of the paper~\cite{pol} is the generalization of the theory of confining strings 
to an arbitrary number of 
colors. Some progress in this direction has been achieved in refs.~\cite{suN1, suN2}. In particular, in ref.~\cite{suN1}, it has been 
shown that at large-$N$
the high-temperature behavior of the free energy of confining strings matches that of QCD.
A related problem is the investigation of the string representation of the 3D Georgi-Glashow
model. This model is different from compact QED due to the presence of the charged W-bosons and due to the 
Higgs-mediated interaction of monopoles which becomes relevant when the Higgs boson is not infinitely heavy.
In this way, the influence of W-bosons to the deconfining phase transition has been analysed in the confining-string language 
in ref.~\cite{kogan}.
String representation of 
the BPS-limit~\cite{bps} (where the mass of the Higgs boson is much smaller than the mass of the 
W-boson) of the $SU(3)$-version of the 3D Georgi-Glashow model has been found in ref.~\cite{nd}. 
In ref.~\cite{mpla}, a representation of the $SU(N)$-version of this model in terms
of monopole densities has been derived, and the potential of these densities has 
been discussed.

The aim of the present paper is to derive and investigate the theory of confining
strings, which would describe the string content of the $SU(N)$-generalization of 3D compact QED, and to generalize it further 
to the 4D-case with the $\theta$-term.
The paper is organized as follows. Sections~2 and~3 are devoted to the 3D-case.
In Section~2, we shall find 
the string tension of the fundamental Wilson loop defined at a flat contour (henceforce called for shortness ``flat Wilson loop''). 
This will be done by the method based on the solution of the system of the saddle-point equations for the dual-photon 
and the magnetic field. [When the magnetic field is eliminated from these equations, the resulting equation for the dual-photon 
field is analogous to the one derived in the usual compact QED in ref.~\cite{polpl}.]
In Section~3, we shall develop the theory of confining strings
based on the Kalb-Ramond field. Using for concreteness the case of fundamental representation, we shall present in Subsection~3.1 
two various methods, by means of which such a theory can be derived. Both of them illustrate how the Kalb-Ramond field 
unifies the monopole and the free-photonic contributions to the Wilson loop.
The string tension derivable in the weak-field limit of this theory (which is nothing else, but
the limit of low monopole densities)
is parametrically the same as the one 
obtained in Section~2 for the flat loop. However, the advantage of the formalism based on the Kalb-Ramond field is that it enables
one not only to fix the proportionality coefficient at the string tension, but also to derive 
the coupling constants of the higher-derivative string terms. 
In Subsection~3.2, the case of the adjoint Wilson loop will be considered, and the respective theory of confining strings
will be constructed in the large-$N$ limit. Section~4 will be devoted to the
generalization of the obtained $SU(N)$-theory of confining strings to the 4D-case with the (field-theoretical) $\theta$-term. As it has been
found in refs.~\cite{cristina, tq} for the usual compact QED, this term
leads to the appearance of the string $\theta$-term.  The latter,
being proportional to the number of self-intersections of the world sheet, might help in
the solution of the problem of crumpling~\cite{fi, dg}~\footnote{This is simply due to the mutual cancellations of contributions to the string
partition function, which originate from any two highly crumpled surfaces, whose numbers of self-intersections differ from each other
by one.}. It will be demonstrated that this indeed happens at (generally) two values of the $\theta$-parameter, which are accessible 
when the electric coupling constant is sufficiently large.

In Summary, the main results of the paper will be discussed.

\section{String tension of the flat Wilson loop in the fundamental representation}

Since the string tension is generated by monopoles, in this Section we shall omit photons. 
The partition function of the grand canonical ensemble of monopoles in the Euclidean space-time reads

\begin{equation}
\label{1}
{\cal Z}_{\rm mon}=\sum\limits_{{\cal N}=0}^{\infty}\frac{\zeta^{\cal N}}{{\cal N}!}
\left<\exp\left[-\frac{g_m^2}{2}\int d^3x d^3y\vec\rho^{\cal N}(\vec x)D_0(\vec x-\vec y)
\vec\rho^{\cal N}(\vec y)\right]\right>_{\rm mon},
\end{equation}
where the monopole density is defined as 
$\vec\rho^{\cal N}(\vec x)=\sum\limits_{k=1}^{\cal N}\vec q_{i_k}\delta(\vec x-\vec z_k)$ at ${\cal N}\ge 1$
and $\vec\rho^{{\cal N}=0}(\vec x)=0$. In these equations, $g_m$ is the magnetic coupling constant of dimensionality $[{\rm mass}]^{-1/2}$, $gg_m=4\pi$,
$\zeta\propto {\rm e}^{-{\rm const}/g^2}$ is the monopole fugacity of dimensionality $[{\rm mass}]^{3}$, 
$D_0(\vec x)=1/(4\pi|\vec x|)$
is the 3D Coulomb propagator, and it has been taken into account that the monopole charges
are distributed along the $(N-1)$-dimensional root vectors $\vec q_i$'s of the group $SU(N)$~\cite{roots}.
Furthermore, the average is defined as follows:

\begin{equation}
\label{monav}
\left<{\cal O}\right>_{\rm mon}=
\prod\limits_{n=0}^{\cal N}\int d^3z_n
\sum\limits_{i_n=\pm 1,\ldots,\pm\frac{N(N-1)}{2}}^{}{\cal O}.
\end{equation}
Upon the explicit summation, the partition function~(\ref{1}) can be represented in the sine-Gordon--type form

\begin{equation}
\label{2}
{\cal Z}_{\rm mon}=
\int {\cal D}\vec\chi\exp\left[-\int d^3x\left(\frac12\left(\partial_\mu\vec\chi\right)^2-2\zeta\sum\limits_{i=1}^{N(N-1)/2}
\cos\left(g_m\vec q_i\vec\chi\right)\right)\right],
\end{equation}
where  $\vec\chi$ is the dual-photon field. By virtue of the formula~\cite{group} 
$\sum\limits_{i=1}^{N(N-1)/2}q_i^\alpha q_i^\beta=\frac{N}{2}\delta^{\alpha\beta}$, $\alpha,\beta=1,\ldots,N-1$,
the following value of the Debye mass of the 
dual photon has been obtained~\cite{mpla}: $m_D=g_m\sqrt{N\zeta}$.

One can further introduce 
the monopole field-strength tensor $\vec F_{\mu\nu}^{\cal N}$ violating the Bianchi identity as $\frac12\varepsilon_{\mu\nu\lambda}
\partial_\mu\vec F_{\nu\lambda}^{\cal N}=g_m\vec\rho^{\cal N}$. Since it was argued to omit photons throughout this Section, we obtain
for $\vec F_{\mu\nu}^{\cal N}$ the following expression:

\begin{equation}
\label{F}
\vec F_{\mu\nu}^{\cal N}\left(\vec x\right)=-g_m\varepsilon_{\mu\nu\lambda}\partial_\lambda\int d^3y D_0\left(\vec x-\vec y\right)
\vec\rho^{\cal N}\left(\vec y\right).
\end{equation}
Then, by virtue of the Stokes' theorem and the formula 

$${\rm tr}{\,}\exp\left(i\vec {\cal O}\vec H\right)=
\sum\limits_{a=1}^{N}\exp\left(i\vec {\cal O}\vec\mu_a\right),$$
where $\vec {\cal O}$ is an arbitrary $(N-1)$-component vector, $\vec H$ is the vector of diagonal generators  
and $\vec\mu_a$'s are the weight vectors of the group $SU(N)$, $a=1,\ldots, N$, we obtain the following expression 
for the Wilson loop defined at the ${\cal N}$-monopole configuration:

\begin{equation}
\label{WcalN}
W(C)_{\rm mon}^{\cal N}=\frac{1}{N}{\,}{\rm tr}{\,}\exp\left(\frac{ig}{2}\vec H\int d^3x\vec F_{\mu\nu}^{\cal N}\Sigma_{\mu\nu}\right)=
\frac{1}{N}\sum\limits_{a=1}^{N}W_a^{\cal N}.
\end{equation}
Here, 

\begin{equation}
\label{Wa}
W_a^{\cal N}\equiv\exp\left(\frac{ig}{2}\vec\mu_a\int d^3x\vec F_{\mu\nu}^{\cal N}\Sigma_{\mu\nu}\right)
\end{equation}
and 

$$\Sigma_{\mu\nu}\left(\vec x\right)=\int\limits_{\Sigma(C)}^{} 
d\sigma_{\mu\nu}\left(\vec x\left(\xi\right)\right)\delta\left(\vec x-\vec x(\xi)\right)$$
is the vorticity tensor current defined at a certain surface 
$\Sigma(C)$ bounded by the contour $C$ and parametrized by the vector 
$\vec x(\xi)$ with $\xi=\left(\xi^1,\xi^2\right)$ standing for the 2D coordinate.
A straightforward calculation with the use of eq.~(\ref{F}) yields

\begin{equation}
\label{W}
W_a^{\cal N}=\exp\left(i\vec\mu_a\int d^3x\vec\rho^{\cal N}\eta\right),
\end{equation}
where 

$$
\eta\left(\vec x; C\right)=\int\limits_{\Sigma(C)}^{} 
d\sigma_\mu\left(\vec x(\xi)\right)\partial_\mu\frac{1}{\left|\vec x-\vec x(\xi)\right|}$$
is the solid angle under which the surface $\Sigma(C)$ is seen by an observer located at the point $\vec x$, $d\sigma_\mu=
\frac12\varepsilon_{\mu\nu\lambda}d\sigma_{\nu\lambda}$. (Note that owing to
the Gauss' law, $\eta\equiv4\pi$ for a closed surface, i.e., when $C$ is shrunk to a point.) The ratio of two $W_a^{\cal N}$'s 
defined at different surfaces, $\Sigma_1$ and $\Sigma_2$, bounded by the contour $C$, reads

$$
\frac{W_a^{\cal N}\left(\Sigma_1\right)}{W_a^{\cal N}\left(\Sigma_2\right)}=\exp\left(i\vec\mu_a
\int\limits_{\Sigma_1\cup\Sigma_2}^{}d\sigma_\mu\left(\vec x(\xi)\right)\int d^3x\vec\rho^{\cal N}
\left(\vec x{\,}\right)\partial_\mu\frac{1}{\left|\vec x-\vec x(\xi)\right|}\right)=$$

\begin{equation}
\label{fract}
=\prod\limits_{k=1}^{\cal N}\exp\left(-i\vec\mu_a\vec q_{i_k}
\int\limits_{\Sigma_1\cup\Sigma_2}^{}d\sigma_\mu\left(\vec x(\xi)\right)\partial_\mu^{\vec x(\xi)}
\frac{1}{\left|\vec z_k-\vec x(\xi)\right|}\right).
\end{equation}
Due to the Gauss' law, the last integral in this equation is equal either to $-4\pi$ or to $0$, depending on whether
the point $\vec z_k$ lies inside or outside the volume bounded by the surface $\Sigma_1\cup\Sigma_2$. Since the product
$\vec\mu_a\vec q_{i_k}$ is equal either to $\pm\frac12$ or to $0$, we conclude that 
$\frac{W_a^{\cal N}\left(\Sigma_1\right)}{W_a^{\cal N}\left(\Sigma_2\right)}=1$. This fact proves the 
independence of $W_a^{\cal N}$ of the choice of the surface $\Sigma$ in the definition~(\ref{Wa}).
Clearly, this is the consequence of the quantization condition $gg_m=4\pi$ employed in the derivation of eq.~(\ref{W}).

A representation of the partition function~(\ref{1}), alternative to eq.~(\ref{2}) and more appropriate for the 
investigation of the Wilson loop, is the one in terms of the 
dynamical monopole densities~\cite{mpla}.
It can be obtained by multiplying eq.~(\ref{1}) by the following unity:

$$
\int {\cal D}\vec\rho\delta\left(\vec\rho-\vec\rho^{\cal N}\right)=
\int {\cal D}\vec\rho{\cal D}\vec\chi\exp\left[ig_m\int d^3x\vec\chi\left(\vec\rho
-\vec\rho^{\cal N}\right)\right],$$
so that the field $\vec\chi$ plays the role of the Lagrange multiplier. We obtain for the partition function:

$$
{\cal Z}_{\rm mon}=\int{\cal D}\vec\rho{\cal D}\vec\chi\exp\left[-\frac{g_m^2}{2}\int d^3x d^3y\vec\rho(\vec x)D_0(\vec x-\vec y)
\vec\rho(\vec y)+ig_m\int d^3x\vec\chi\vec\rho\right]\times$$

\begin{equation}
\label{II}
\times\sum\limits_{{\cal N}=0}^{\infty}\frac{\zeta^{\cal N}}{{\cal N}!}
\left<\exp\left(-ig_m\int d^3x\vec\chi\vec\rho^{\cal N}\right)\right>_{\rm mon},
\end{equation}
where the last sum is equal to 

\begin{equation}
\label{III}
\exp\left[2\zeta\int d^3x
\sum\limits_{i=1}^{N(N-1)/2}
\cos\left(g_m\vec q_i\vec\chi\right)\right].
\end{equation}
Accordingly, eq.~(\ref{W}) becomes replaced by

$$
W_a^{\cal N}\to W_a=\frac{1}{{\cal Z}_{\rm mon}}
\int{\cal D}\vec\rho{\cal D}\vec\chi\exp\Biggl\{-\frac{g_m^2}{2}\int d^3x d^3y\vec\rho(\vec x)D_0(\vec x-\vec y)
\vec\rho(\vec y)+$$

$$+\int d^3x\Biggl[ig_m\vec\chi\vec\rho+
2\zeta
\sum\limits_{i=1}^{N(N-1)/2}
\cos\left(g_m\vec q_i\vec\chi\right)+
i\vec\mu_a\vec\rho\eta\Biggr]\Biggr\},$$
and the full expression for the monopole contribution to the Wilson loop [instead of eq.~(\ref{WcalN})]
reads $W(C)_{\rm mon}=\frac{1}{N}\sum\limits_{a=1}^{N}W_a$.

Let us next introduce the magnetic field according to the formulae 

$$\partial_\mu\vec B_\mu=\vec\rho,~~ 
\varepsilon_{\mu\nu\lambda}\partial_\nu\vec B_\lambda=0.$$ 
This yields

$$
W_a=\frac{1}{{\cal Z}_{\rm mon}}
\int{\cal D}\vec B_\mu\delta\left(\varepsilon_{\mu\nu\lambda}\partial_\nu\vec B_\lambda\right)\times$$

\begin{equation}
\label{WWW}
\times\int {\cal D}\vec\chi
\exp\Biggl\{\int d^3x\Biggl[-\frac{g_m^2}{2}\vec B_\mu^2+
ig_m\vec\chi\partial_\mu\vec B_\mu+
2\zeta
\sum\limits_{i=1}^{N(N-1)/2}
\cos\left(g_m\vec q_i\vec\chi\right)\Biggr]+
4\pi i\vec\mu_a\int\limits_{\Sigma(C)}^{}d\sigma_\mu\vec B_\mu\Biggr\},
\end{equation}
where, as it has been discussed, the surface $\Sigma(C)$ is arbitrary. 
The magnetic and the dual-photon fields 
can be integrated out of eq.~(\ref{WWW}) by solving the respective saddle-point equations. 
From now on in this Section, we shall 
consider the contour $C$ located in the $(x,y)$-plane. This naturally leads to the following Ans\"atze for the  
fields to be inserted into the saddle-point equations: 
$\vec B_\mu=\delta_{\mu 3}\vec B(z)$, $\vec\chi=\vec\chi(z)$. 
For the points 
$(x,y)$ lying inside the contour $C$, these equations then read:

\begin{equation}
\label{sp1}
ig_m\vec\chi'+g_m^2\vec B-4\pi i\vec\mu_a\delta(z)=0,
\end{equation}

\begin{equation}
\label{sp2}
i\vec B'-2\zeta\sum\limits_{i=1}^{N(N-1)/2}\vec q_i\sin\left(g_m\vec q_i\vec\chi\right)=0,
\end{equation}
where $'\equiv d/dz$.
[Note that differentiating eq.~(\ref{sp1}) and substituting the result into eq.~(\ref{sp2}), we obtain the equation

$$\vec\chi''-2g_m\zeta\sum\limits_{i=1}^{N(N-1)/2}\vec q_i\sin\left(g_m\vec q_i\vec\chi\right)=g\vec\mu_a
\delta'(z),$$
which is the $SU(N)$-generalization of the respective saddle-point equation obtained in ref.~\cite{polpl}.]
These equations can be solved by adopting one extra Ansatz for the saddle-points of the fields, 
$\vec B(z)=\vec\mu_a B(z)$, $\vec\chi(z)=\vec\mu_a\chi(z)$, that due to the formula 

\begin{equation}
\label{norm}
\vec\mu_a\vec\mu_b=\frac12\left(\delta_{ab}-\frac{1}{N}\right)
\end{equation}
makes $W_a$ $a$-independent. Let us next multiply eq.~(\ref{sp2}) by $\vec\mu_a$, taking into account that
for any $a$, $(N-1)$ positive roots yield the scalar product with $\vec\mu_a$ equal to $1/2$, while the others are 
orthogonal to $\vec\mu_a$. Equations~(\ref{sp1}), (\ref{sp2}) then go over to

\begin{equation}
\label{SP}
2i\psi'+g_m^2B=4\pi i\delta(z),~~ B'+2i\zeta N\sin\psi=0,
\end{equation}
where $\psi\equiv g_m\chi/2$. One can straightforwardly check that the solution to this system of equations has the following form:

\begin{equation}
\label{solution}
B(z)=i\frac{8m_D}{g_m^2}\frac{{\rm e}^{-m_D|z|}}{1+{\rm e}^{-2m_D|z|}},~~
\psi(z)=4{\,}{\rm sgn}{\,}z\cdot\arctan\left({\rm e}^{-m_D|z|}\right).
\end{equation}
Taking the value $B(0)$ for the evaluation of the Wilson loop, we obtain for the string tension:

\begin{equation}
\label{Sig}
\sigma=4\pi\cdot\frac{N-1}{2N}\frac{4m_D}{g_m^2}=
\frac{8\pi}{g_m}\frac{N-1}{\sqrt{N}}\sqrt{\zeta}.
\end{equation}
This result demonstrates explicitly the nonanalytic behavior of $\sigma$ with $g$, but 
does not yield the correct overall numerical factor. That is because $B(z)$, necessary for this calculation, is defined ambiguously. 
The ambiguity originates from the exponentially
large thickness of the string: $|z|<{\cal O}\left(m_D^{-1}\right)$. It is therefore unclear what value of $B(z)$
we should take: either $B(0)$, as it was done, or the one averaged over some range of $z$. It turns out that in the weak-field (low-density) limit, 
the problem can be solved for an arbitrarily-shaped surface, by using the representation in terms of the Kalb-Ramond field, which will
be described in the next Section.

\section{SU(N) confining strings}

\subsection{Fundamental representation}

One of the ways to introduce 
the Kalb-Ramond field is to do it first for the description 
of the dual photon, by considering this field as the field-strength 
tensor corresponding to the field $\vec B_\mu$, namely $\vec B_\mu=\frac{1}{2g_m}\varepsilon_{\mu\nu\lambda}\vec h_{\nu\lambda}$.
The Wilson loop~(\ref{WWW}) then takes the form~\footnote{Note the following correspondence between the 
Coulomb interaction of monopole densities and the actions of the magnetic- and Kalb-Ramond fields:

$$
\frac14\int d^3x\vec h_{\mu\nu}^2=\frac{g_m^2}{2}\int d^3x\vec B_\mu^2=
\frac{g_m^2}{2}\int d^3x d^3y\vec\rho(\vec x)D_0(\vec x-\vec y)
\vec\rho(\vec y).$$}

$$
W_a=
\frac{1}{{\cal Z}_{\rm mon}}
\int{\cal D}\vec h_{\mu\nu}\delta\left(\partial_\mu\vec h_{\mu\nu}\right)\int {\cal D}\vec\chi
\exp\Biggl\{\int d^3x\Biggl[-\frac14\vec h_{\mu\nu}^2+$$

\begin{equation}
\label{WwW}
+\frac{i}{2}\vec\chi\varepsilon_{\mu\nu\lambda}\partial_\mu\vec h_{\nu\lambda}+
2\zeta
\sum\limits_{i=1}^{N(N-1)/2}
\cos\left(g_m\vec q_i\vec\chi\right)\Biggr]+
\frac{ig}{2}\vec\mu_a\int\limits_{\Sigma(C)}^{}d\sigma_{\mu\nu}\vec h_{\mu\nu}\Biggr\}.
\end{equation}
The field $\vec\chi$ can be further integrated out by solving the saddle-point equation of the form~(\ref{sp2}),
where $\vec B'$ is replaced by  
$\frac{1}{2g_m}\varepsilon_{\mu\nu\lambda}\partial_\mu\vec h_{\nu\lambda}$ (and $\vec\chi$ depends on all three coordinates). 
Using the Ansatz $\vec h_{\mu\nu}=\vec\mu_a 
h_{\mu\nu}$, we arrive at the following substitution in eq.~(\ref{WwW}):

$$
\int d^3x\left[\frac{i}{2}\vec\chi\varepsilon_{\mu\nu\lambda}\partial_\mu\vec h_{\nu\lambda}+
2\zeta
\sum\limits_{i=1}^{N(N-1)/2}
\cos\left(g_m\vec q_i\vec\chi\right)\right]\Longrightarrow$$

\begin{equation}
\label{Vh}
\Longrightarrow -V\left[\vec h_{\mu\nu}\right]\equiv -2(N-1)\zeta\int d^3x \left[H_a{\rm arcsinh}{\,} H_a
-\sqrt{1+H_a^2}\right].
\end{equation}
Here, 

\begin{equation}
\label{Ha}
H_a\equiv \frac{1}{2g_m\zeta (N-1)}\vec\mu_a\varepsilon_{\mu\nu\lambda}\partial_\mu\vec h_{\nu\lambda},
\end{equation} 
and $V\left[\vec h_{\mu\nu}\right]$ is the 
multivalued potential of the Kalb-Ramond field (or of monopole densities).
Similarly to the case of compact QED~\cite{pol}, it is the summation over branches of this potential, which yields the 
summation over world sheets $\Sigma(C)$ in eq.~(\ref{WwW}). 

Another way to discuss the connection of the Kalb-Ramond field with the string world sheets is based on the 
semi-classical analysis of the saddle-point equations analogous to eqs.~(\ref{SP}):

\begin{equation}
\label{SSP1}
2i\partial_\mu\psi+\frac{g_m}{2}\varepsilon_{\mu\nu\lambda}h_{\nu\lambda}=4\pi i\Sigma_\mu,
\end{equation}

\begin{equation}
\label{SSP2}
\frac{1}{2g_m}\varepsilon_{\mu\nu\lambda}\partial_\mu h_{\nu\lambda}+2i\zeta N\sin\psi=0.
\end{equation}
Here, $\Sigma_\mu\equiv\frac12\varepsilon_{\mu\nu\lambda}\Sigma_{\nu\lambda}$,
and the field $\psi$, by means of which the correspondence between the Kalb-Ramond field and the stationary surface is established,
is defined by the relation $\vec\chi=\frac{2}{g_m}\vec\mu_a\psi$. In this approach, based on the (auxiliary) field $\psi$,
the summation over branches of the potential $V\left[\vec h_{\mu\nu}\right]$ is equivalent to the following procedure.
One should restrict oneself to the domain $|\psi|\le\pi$ and solve eqs.~(\ref{SSP1}), (\ref{SSP2}) with the conditions

\begin{equation}
\label{cond1}
\lim\limits_{|\vec x|\to\infty}^{}\psi(\vec x)=0,
\end{equation}

\begin{equation}
\label{cond2}
\lim\limits_{\varepsilon\to 0}^{}\left\{\psi\left[\vec x(\xi)+\varepsilon\vec n(\xi)\right]-
\psi\left[\vec x(\xi)-\varepsilon\vec n(\xi)\right]\right\}=2\pi,
\end{equation}
where $\vec n(\xi)$ is the normal vector to $\Sigma$ at an arbitrary point $\vec x(\xi)$. After that, one should 
get rid of the so-appearing $\Sigma$-dependence of $W_a$ by extremizing the latter with respect to $\vec x(\xi)$.
Owing to the formula 

$$
\delta\int d\sigma_{\mu\nu}h_{\mu\nu}\left[\vec x(\xi)\right]=\int d\sigma_{\mu\nu}H_{\mu\nu\lambda}\left[\vec x(\xi)\right]\delta x_\lambda(\xi),
$$
such an extremization is equivalent to the definition of the extremal surface through the equation 
$H_{\mu\nu\lambda}\left[\vec x(\xi)\right]=0$. By virtue of eqs.~(\ref{SSP2}), (\ref{cond2}), $H_{\mu\nu\lambda}$ 
indeed vanishes on both sides of the stationary surface. The reason for that is the following 
distinguished property of the extremal surface:
$\psi$ is equal to $\pi$ on one of its sides and to $-\pi$ on the other, whereas for any other surface, 
$|\psi|$ is smaller than $\pi$ on one side and larger than $\pi$ on the other side. Therefore, 
according to eq.~(\ref{SSP2}), at both
sides of the extremal surface, $\varepsilon_{\mu\nu\lambda}\partial_\mu h_{\nu\lambda}=0$. This is 
equivalent to the 
equation $\varepsilon_{\mu\nu\lambda}H_{\mu\nu\lambda}=0$ and hence (upon the multiplication of both sides by $\varepsilon_{\mu'\nu'\lambda'}$)
to $H_{\mu\nu\lambda}=0$.
[In particular, for a flat contour the extremal surface is the flat surface inside this contour.
This can explicitly be seen from the solutions~(\ref{solution}) of eqs.~(\ref{SP}). Namely, at any $z$ such that $|z|\ll m_D^{-1}$,
$\psi(z)\simeq\pi\cdot {\rm sgn} z$, $B'(z)\simeq 0$, where both equalities hold with the exponential accuracy. The equation $B'(z)\simeq 0$
is equivalent to the condition of vanishing of $H_{\mu\nu\lambda}$ at both sides of the flat surface.]

According to the Hodge decomposition theorem, the general form of $\vec h_{\mu\nu}$ is 
$\vec h_{\mu\nu}=\partial_\mu\vec A_\nu-\partial_\nu\vec A_\mu+\varepsilon_{\mu\nu\lambda}\partial_\lambda\vec\phi$.
The constraint $\partial_\mu\vec h_{\mu\nu}=0$, imposed in eq.~(\ref{WwW}), is equivalent to setting $\vec A_\mu$ equal zero.
From now on, we shall promote $\vec h_{\mu\nu}$ to include also the fields of free photons, $\vec A_\mu$, by abolishing this 
constraint. Let us further go into the weak-field limit, $|H_a|\ll 1$. Using the Cauchy inequality, we have 

\begin{equation}
\label{wf}
\left|H_a\right|\le\frac{\left|\vec\mu_a\right|\left|\vec\rho
\right|}{\zeta(N-1)}=
\frac{|\vec\rho|}{\zeta\sqrt{2N(N-1)}},
\end{equation}
so that the weak-field limit is equivalent to the following low-density approximation: $|\vec\rho|\ll\zeta\sqrt{2N(N-1)}$.
To understand in what sense this inequality implies the low-density approximation, note that the mean monopole density stemming from eq.~(\ref{2}) reads:

\begin{equation}
\label{me}
\left|\vec\rho\right|_{\rm mean}=\frac{\partial\ln{\cal Z}_{\rm mon}}{{\cal V}\partial\ln\zeta}\simeq\zeta N(N-1),
\end{equation}
where ${\cal V}$ is the three-volume occupied by the system.
Therefore, at large-$N$, the low-density approximation implies that $|\vec\rho|$ should be of the order $N$ times smaller than its mean value.

In the weak-field limit, we now have the following expression for the total Wilson loop, where the constraint 
$\partial_\mu\vec h_{\mu\nu}=0$ is removed:

\begin{equation}
\label{Wwf}
W\left(C,\Sigma\right)_{\rm weak-field}^{\rm tot}=\frac{1}{{\cal Z}^{\rm tot}}
\int{\cal D}\vec h_{\mu\nu}\exp\left\{-\int d^3x\left[\frac{1}{12m_D^2}\vec H_{\mu\nu\lambda}^2
+\frac14\vec h_{\mu\nu}^2-\frac{ig}{2}\vec\mu_a\vec h_{\mu\nu}\Sigma_{\mu\nu}\right]\right\}.
\end{equation}
Here, ${\cal Z}^{\rm tot}$ is given by the same integral over $\vec h_{\mu\nu}$-field, but with $\Sigma_{\mu\nu}$ set to zero, and 
$\vec H_{\mu\nu\lambda}=\partial_\mu\vec h_{\nu\lambda}+\partial_\lambda\vec h_{\mu\nu}+
\partial_\nu\vec h_{\lambda\mu}$ is the Kalb-Ramond field-strength tensor. [The apparent $\Sigma$-dependence of the r.h.s. of eq.~(\ref{Wwf})
is due to the fact that in course of the $\vec h_{\mu\nu}$-expansion of $V\left[\vec h_{\mu\nu}\right]$, only one branch of this potential
has been taken into account. As it was discussed above, the $\Sigma$-dependence disappears upon the summation over all the branches.]
Owing to the Hodge decomposition theorem for $\vec h_{\mu\nu}$,
eq.~(\ref{Wwf}) is obviously factorized as $W\left(C,\Sigma\right)_{\rm weak-field}^{\rm tot}=W_a W_a^{\rm phot}$, where the free 
photonic contribution to the Wilson loop reads 

\begin{equation}
\label{Fp}
W_a^{\rm phot}=\exp\left[-g^2\frac{N-1}{4N}\oint\limits_{C}^{}dx_\mu\oint\limits_{C}^{}dx'_\mu D_0\left(\vec x-\vec x'\right)\right].
\end{equation}
Doing the integration over $\vec h_{\mu\nu}$ in eq.~(\ref{Wwf}), we obtain

$$W\left(C,\Sigma\right)_{\rm weak-field}^{\rm tot}=
\exp\Biggl\{-g^2\frac{N-1}{4N}\Biggl[\oint\limits_C^{}dx_\mu\oint\limits_C^{}dx'_\mu
D_{m_D}(\vec x-\vec x')+$$

\begin{equation}
\label{WCSig}
+\frac{m_D^2}{2}\int\limits_{\Sigma(C)}^{}d\sigma_{\mu\nu}(\vec x(\xi))
\int\limits_{\Sigma(C)}^{}d\sigma_{\mu\nu}(\vec x(\xi'))D_{m_D}\left(\vec x(\xi)-\vec x(\xi')\right)
\Biggr]\Biggr\},
\end{equation}
where $D_{m_D}\left(\vec x\right)={\rm e}^{-m_D|\vec x|}/(4\pi|\vec x|)$ is the 3D Yukawa propagator.
Note that the free photonic contribution is completely canceled out of this expression, i.e., it is only the dual photon 
(of the mass $m_D$) which mediates the $C\times C$- and $\Sigma\times\Sigma$-interactions.

One can further expand the nonlocal string effective action 

\begin{equation}
\label{snonloc}
S_{\rm str}=(gm_D)^2\frac{N-1}{8N}\int\limits_{\Sigma(C)}^{}d\sigma_{\mu\nu}(\vec x(\xi))
\int\limits_{\Sigma(C)}^{}d\sigma_{\mu\nu}(\vec x(\xi'))D_{m_D}\left(\vec x(\xi)-\vec x(\xi')\right)
\end{equation}
in powers of the derivatives with respect to the world-sheet coordinates $\xi^a$'s. Note that the actual parameter of this expansion 
is $1/(m_DR)^2$, where $R\sim\sqrt{{\rm Area}(\Sigma)}$ is the size of $\Sigma$ (see the discussion below). The resulting quasi-local action reads
(cf. refs.~\cite{dva, cristina}):

\begin{equation}
\label{sstr}
S_{\rm str}=\sigma\int d^2\xi\sqrt{\sf g}+
\alpha^{-1}\int d^2\xi\sqrt{\sf g}{\sf g}^{ab}
(\partial_at_{\mu\nu})(\partial_bt_{\mu\nu})+\kappa
\int d^2\xi\sqrt{\sf g}{\cal R}+{\cal O}\left(\frac{\sigma}{m_D^4R^2}\right).
\end{equation}
Here, 
$\partial_a\equiv\partial/\partial\xi^a$, and the following quantities characterize $\Sigma$: ${\sf g}_{ab}(\xi)=
(\partial_a x_\mu(\xi))(\partial_b x_\mu(\xi))$ is the induced-metric tensor, ${\sf g}=\det\| {\sf g}^{ab}\|$,
$t_{\mu\nu}(\xi)=\varepsilon^{ab}(\partial_a 
x_\mu(\xi))(\partial_b x_\nu(\xi))/\sqrt{\sf g}$ is the extrinsic-curvature tensor,
${\cal R}=\left(\partial^a\partial_a\ln
\sqrt{\sf g}\right)/\sqrt{\sf g}$ is the (conformal-gauge expression for the) scalar (or intrinsic) curvature.

The third 
term on the r.h.s. of eq.~(\ref{sstr})
is known to be a full derivative, and therefore it does not actually contribute to the string effective action, while 
the second 
term describes the so-called rigidity of the string~\cite{fi}.
The reason for the notation $\alpha^{-1}$, introduced in ref.~\cite{fi}, is that, as it has been shown in that paper, it is $\alpha$, 
which is asymptotically free. This asymptotic freedom then indicates that the rigidity term can only be IR relevant, provided the respective $\beta$-function
has a zero in the IR region. However, since the time of the paper~\cite{fi}, such a zero has not been
found. This fact makes unclear a possible relevance of the rigidity term to the solution of the old-standing 
problem of crumpling of large world sheets in the Euclidean space-time and necessitates to seek other possible solutions of this problem.
One of these, based on the string $\theta$-term, has been proposed already in ref.~\cite{fi}. A possible derivation of such a term, within the 
model under discussion, will be presented in Section~4.

The string coupling constants in eq.~(\ref{sstr}) read

\begin{equation}
\label{tef}
\sigma=2\pi^2\frac{N-1}{\sqrt{N}}\frac{\sqrt{\zeta}}{g_m},
\end{equation}

\begin{equation}
\label{alf}
\alpha^{-1}=-3\kappa/2=-\frac{\pi^2(N-1)}{4g_m^3N^{3/2}\sqrt{\zeta}}.
\end{equation} 
In the weak-field (or low-density) approximation, we can therefore fix the 
proportionality factor $2\pi^2$ at $\sigma$, that is close to the factor $8\pi$ of eq.~(\ref{Sig}) which, however, could have not 
been fixed within the method of Section~2. Another important fact is that 
(in the weak-field approximation) the string tension and higher string coupling constants 
are the same for all large enough surfaces $\Sigma(C)$. [The words ``large enough'' here mean the validity of the inequality 
$m_DR\gg 1$, i.e., the string length $R$ should significantly exceed the exponentially large string thickness $m_D^{-1}$. Indeed, 
the general $n$-th term of the derivative (or curvature) expansion of $S_{\rm str}$ is of the order of $\sigma m_D^2R^4(m_DR)^{-2n}$.]
It is, however, worth noting that some of the terms in the expansion~(\ref{sstr}) vanish at the surface of the minimal area corresponding to
a given contour $C$ (e.g., at the flat surface, in the case considered in Section~2). This concerns, for instance, the rigidity term~\footnote{
Indeed, one can prove the equality 
${\sf g}^{ab}(\partial_at_{\mu\nu})(\partial_bt_{\mu\nu})=(D^aD_ax_\mu)(D^bD_bx_\mu)$, where 
$D_aD_bx_\mu=\partial_a\partial_bx_\mu-\Gamma_{ab}^c\partial_cx_\mu$, and $\Gamma_{ab}^c$ is the Christoffel symbol corresponding to 
the metric ${\sf g}^{ab}$. On the other hand, the surface of the minimal area is defined by the equation $D^aD_ax_\mu(\xi)=0$ together 
with the respective boundary condition at the contour $C$.}.

Let us finally demonstrate how to introduce the Kalb-Ramond field in the way alternative to its definition via the $\vec B_\mu$-field 
and incorporating automatically the free-photonic contribution to the Wilson loop. It is based on the 
direct application of eqs.~(\ref{1}) and (\ref{W}), that yields

$$
W_a=\sum\limits_{{\cal N}=0}^{\infty}\frac{\zeta^{\cal N}}{{\cal N}!}
\left<\exp\left[-\frac{g_m^2}{2}\int d^3x d^3y\vec\rho^{\cal N}(\vec x)D_0(\vec x-\vec y)
\vec\rho^{\cal N}(\vec y)+i\vec\mu_a\int d^3x\vec\rho^{\cal N}\eta\right]\right>_{\rm mon}=$$

\begin{equation}
\label{alphaa}
=\frac{1}{{\cal Z}_{\rm mon}}\int {\cal D}\vec\varphi\exp\Biggl\{-\int d^3x\Biggl[\frac12\left(\partial_\mu\vec\varphi-\frac{\vec\mu_a}{g_m}
\partial_\mu\eta\right)^2-2\zeta\sum\limits_{i=1}^{N(N-1)/2}
\cos\left(g_m\vec q_i\vec\varphi\right)\Biggr]\Biggr\},
\end{equation}
where $\vec\varphi=\vec\chi+\frac{\vec\mu_a}{g_m}\eta$. One can further use the following formula which 
shows explicitly how the Kalb-Ramond field unifies the monopole and the free-photonic contributions
to the Wilson loop:

$$\exp\left[-\frac12\int d^3x\left(\partial_\mu\vec\varphi-\frac{\vec\mu_a}{g_m}
\partial_\mu\eta\right)^2-g^2\frac{N-1}{4N}
\oint\limits_C^{}dx_\mu\oint\limits_C^{}dx'_\mu
D_0(\vec x-\vec x')\right]=$$

\begin{equation}
\label{unify}
=\int {\cal D}\vec h_{\mu\nu}\exp\left[-\int d^3x\left(
\frac14\vec h_{\mu\nu}^2+\frac{i}{2}\vec\varphi\varepsilon_{\mu\nu\lambda}\partial_\mu
\vec h_{\nu\lambda}-\frac{ig}{2}\vec\mu_a\vec h_{\mu\nu}\Sigma_{\mu\nu}\right)\right].
\end{equation}
In fact, one can prove that both sides of this formula are equal to~\footnote{
Obviously, in the noncompact case, when monopoles are disregarded and $\vec h_{\mu\nu}=\partial_\mu\vec A_\nu-
\partial_\nu\vec A_\mu$, the r.h.s. of eq.~(\ref{unify}) yields the free-photonic contribution to the Wilson loop, 
eq.~(\ref{Fp}).} 

$$\exp\left[-\frac12\int d^3x\left(\partial_\mu\vec\varphi+g\vec\mu_a\Sigma_\mu\right)^2\right].$$
Inserting further eq.~(\ref{unify}) into eq.~(\ref{alphaa}), we obtain

$$W_a=\frac{1}{{\cal Z}^{\rm tot}}
\int{\cal D}\vec h_{\mu\nu}{\cal D}\vec\varphi\exp\Biggl[-\int d^3x\Biggl(
\frac14\vec h_{\mu\nu}^2+$$

$$
+\frac{i}{2}\vec\varphi\varepsilon_{\mu\nu\lambda}\partial_\mu
\vec h_{\nu\lambda}-2\zeta\sum\limits_{i=1}^{N(N-1)/2}\cos\left(g_m\vec q_i\vec\varphi\right)
-\frac{ig}{2}\vec\mu_a\vec h_{\mu\nu}\Sigma_{\mu\nu}\Biggr)\Biggr].$$
Denoting $\vec\chi\equiv-\vec\varphi$, 
we arrive back to eq.~(\ref{WwW}) with the constraint $\partial_\mu\vec h_{\mu\nu}=0$ abolished, as it should be.

\subsection{Adjoint representation}

Let us now extend the ideas of the previous Subsection to the case of the Wilson loop in the adjoint representation. 
The charges of quarks in this representation are distributed along the 
roots, so that in the formulae for the Wilson loop, eqs.~(\ref{Wa}) and (\ref{W}), $\vec\mu_a$ should be replaced by $\vec q_i$. Noting
that any root is a difference of two weights, we shall henceforth in this Section label roots by two indices
running from $1$ to $N$, e.g., $\vec q_{ab}=\vec\mu_a-\vec\mu_b$~\footnote{In particular, this makes explicit 
the number of positive roots, $\frac{N^2-N}{2}$.}. Equation~(\ref{norm}) then leads to the following formula:

\begin{equation}
\label{qq}
\vec q_{ab}\vec q_{cd}=\frac12\left(\delta_{ac}+\delta_{bd}-\delta_{ad}-\delta_{bc}\right),
\end{equation}
according to which the product $\vec q_{ab}\vec q_{cd}$ may take the values $0, \pm \frac12, \pm 1$. Owing to this fact,
the ratio~(\ref{fract}) is again equal to 1, i.e., for adjoint quarks, whose charge obeys the quantization condition 
$gg_m=4\pi$, the Wilson loop is as surface independent, as it is for fundamental quarks.

The adjoint-case version of eq.~(\ref{WwW}) (with the constraint $\partial_\mu\vec h_{\mu\nu}=0$ abolished) has the form:

$$
W_{ab}=
\frac{1}{{\cal Z}^{\rm tot}}
\int{\cal D}\vec h_{\mu\nu}{\cal D}\vec\chi
\exp\Biggl\{\int d^3x\Biggl[-\frac14\vec h_{\mu\nu}^2+$$

\begin{equation}
\label{wab}
+\frac{i}{2}\vec\chi\varepsilon_{\mu\nu\lambda}\partial_\mu\vec h_{\nu\lambda}+
\zeta\sum\limits_{c,d=1}^{N}
\cos\left(g_m\vec q_{cd}\vec\chi\right)\Biggr]+
\frac{ig}{2}\vec q_{ab}\int\limits_{\Sigma(C)}^{}d\sigma_{\mu\nu}\vec h_{\mu\nu}\Biggr\}.
\end{equation}
The saddle-point equation, emerging from the above formula in the course of integration over $\vec\chi$,

\begin{equation}
\label{neweq}
\frac{i}{2}\vec\chi\varepsilon_{\mu\nu\lambda}\partial_\mu\vec h_{\nu\lambda}=
g_m\zeta\sum\limits_{c,d=1}^{N}\vec q_{cd}
\sin\left(g_m\vec q_{cd}\vec\chi\right),
\end{equation}
is to be solved in a way similar to the one of the previous Subsection, namely by using the Ansatz
$\vec h_{\mu\nu}=\vec q_{ab}h_{\mu\nu}$, $\vec\chi=\vec q_{ab}\chi$. Multiplying both sides of eq.~(\ref{neweq})
by $\vec q_{ab}$, we obtain:

\begin{equation}
\label{rhs}
\frac{i}{2g_m\zeta}\varepsilon_{\mu\nu\lambda}\partial_\mu h_{\nu\lambda}=
\sum\limits_{c,d=1}^{N}\vec q_{ab}\vec q_{cd}\sin\left(g_m\vec q_{ab}\vec q_{cd}\chi\right).
\end{equation}
Let us now compute the sum on the r.h.s. of this equation by virtue of eq.~(\ref{qq}).
To this end, imagine ourselves the antisymmetric $N\times N$-matrix of roots $\vec q_{cd}$'s.
Obviously, for a fixed root $\vec q_{ab}$, there is one root $\vec q_{cd}$ (equal to $\vec q_{ab}$) and 
a negative symmetric to it, whose
scalar products with $\vec q_{ab}$ are equal to $1$ and $-1$. Our aim is to calculate the number $n$ of roots,
whose scalar product with $\vec q_{ab}$ is equal to $\pm\frac12$~\footnote{For all other $N^2-N-2-n$ roots, the scalar product
vanishes, and so does the r.h.s. of eq.~(\ref{rhs}).}. According to eq.~(\ref{qq}), these roots belong either to the rows
$c=a$, $c=b$, or to the columns $d=a$, $d=b$, which in total contain $4N-4$ elements. Among these, two are diagonal and 
other two are those, which yield the scalar product $\pm 1$. The remaining $n=4(N-2)$ roots are precisely that, which 
yield the scalar product $\pm\frac12$, so that this product equals $\frac12$ for $2(N-2)$ of them and $-\frac12$ 
for the other $2(N-2)$. The sum on the r.h.s. of eq.~(\ref{rhs}) thus takes the form

$$
\sin(2\psi)-\sin(-2\psi)+2(N-2)\left[\frac12\sin\psi-\frac12\sin(-\psi)\right]=
2\left[\sin(2\psi)+(N-2)\sin\psi\right],
$$
where the field $\psi$ has been defined after eq.~(\ref{SP}). Let us now adopt the limit $N\gg 1$, in which eq.~(\ref{rhs})
takes the form $\sin\psi=iH_{ab}$, where [cf. eq.~(\ref{Ha})]

$$H_{ab}\equiv\frac{1}{4Ng_m\zeta}\vec q_{ab}\varepsilon_{\mu\nu\lambda}\partial_\mu\vec h_{\nu\lambda}.$$
The analogue of the potential~(\ref{Vh}) in the same limit then reads

$$
V\left[\vec h_{\mu\nu}\right]=4N\zeta\int d^3x \left[H_{ab}{\,}{\rm arcsinh}{\,} H_{ab}
-\sqrt{1+H_{ab}^2}\right].$$
The string representation of the adjoint Wilson loop in the large-$N$ limit is therefore given by eq.~(\ref{wab}) with the 
substitution 

$$
\int d^3x\left[\frac{i}{2}\vec\chi\varepsilon_{\mu\nu\lambda}\partial_\mu\vec h_{\nu\lambda}+
\zeta\sum\limits_{c,d=1}^{N}
\cos\left(g_m\vec q_{cd}\vec\chi\right)\right]\Longrightarrow -V\left[\vec h_{\mu\nu}\right].$$

We are finally interested in the adjoint-case counterparts of eqs.~(\ref{tef}), (\ref{alf}), one can obtain in the 
weak-field limit $\left|H_{ab}\right|\ll 1$~\footnote{Note that, using the formula $H_{ab}=\frac{1}{2N\zeta}\vec q_{ab}\vec\rho$
and the Cauchy inequality, one gets the following analogue of eq.~(\ref{wf}): 
$\left|H_{ab}\right|\le\frac{\left|\vec\rho\right|}{2N\zeta}$. This clearly leads to the same definition of the 
weak-field limit in terms of the low-density approximation, as in the fundamental case. Namely, the weak-field limit
corresponds to densities $\left|\vec\rho\right|$, which are of the order $N$ times smaller than the mean one~(\ref{me}).}.
In this limit, the formula~(\ref{Wwf}) recovers itself, with the substitution $\vec\mu_a\to\vec q_{ab}$. As a consequence, the ratios of adjoint-case
values of string couplings $\sigma$, $\alpha^{-1}$, and $\kappa$ to the respective fundamental-case values, eqs.~(\ref{tef}) and (\ref{alf}),
is equal to $\frac{2N}{N-1}\simeq2$. In particular, for the string tensions this ratio coincides with the known 
leading large-$N$ QCD-result (see e.g. ref.~\cite{mak}).

\section{Generalization to the 4D-case with the $\theta$-term}

In this Section, we shall consider the 4D-case and introduce the field-theoretical $\theta$-term~\footnote{In the case when the ensemble of 
Abelian-projected monopoles is modeled by the magnetically-charged dual Higgs field, the effects, produced by this term to the respective 
string effective action, have been studied in ref.~\cite{term}.}.
As it has been first found for compact QED in refs.~\cite{cristina, tq}, by means of the derivative expansion of the resulting 
nonlocal string effective action,
this term generates the string $\theta$-term. Being proportional to the number of self-intersections of the world sheet,
the latter might be important for the solution of the problem of crumpling of large world sheets~\cite{dg, fi}.
In this Section, we shall perform the respective analysis for the general $SU(N)$-case under study, both in the fundamental and 
in the adjoint representations.
It is worth noting that in the lattice 4D compact QED, confinement holds only at strong coupling. On the opposite,
the continuum counterpart~\cite{cristina, tq} (of the $SU(N)$-version) of this model, we are going to explore,
possesses confinement at arbitrary values of coupling. However, we shall see that the solution to the problem of crumpling
due to the $\theta$-term can only be possible in the strong-coupling regime, implied in a certain sense.
Note also that the continuum
sine-Gordon theory of the dual-photon field possesses an UV cutoff, which appears in course of the path-integral average over the shapes of
monopole loops.
However, the $\theta$-parameter is dimensionless, and consequently its values,
at which one may expect the disappearance of crumpling, will be cutoff-independent.

The full partition function, including the $\theta$-term and the average over free photons, reads [cf. eq.~(\ref{1})]

\begin{equation}
\label{Z4D}
{\cal Z}=\int {\cal D}\vec A_\mu{\rm e}^{-\frac14\int d^4x\vec F_{\mu\nu}^2}\sum\limits_{{\cal N}=0}^{\infty}
\frac{\zeta^{\cal N}}{{\cal N}!}\left<\exp\left\{-{\cal S}\left[\vec j_\mu^{\cal N}, \vec A_\mu\right]\right\}\right>_{\rm mon},
\end{equation}
where

\begin{equation}
\label{Sini}
{\cal S}\left[\vec j_\mu^{\cal N}, \vec A_\mu\right]=
\frac{1}{2}\int d^4xd^4y\vec j_\mu^{\cal N}(x)
D_0(x-y)\vec j_\mu^{\cal N}(y)+
\frac{i\theta g^2}{8\pi^2}
\int d^4x\vec A_\mu\vec j_\mu^{\cal N},
\end{equation}
and $\vec F_{\mu\nu}=\partial_\mu\vec A_\nu-\partial_\nu\vec A_\mu$.
For ${\cal N}=0$, the monopole current $j_\mu^{\cal N}$ is equal to zero, whereas for
${\cal N}\ge 1$, it is defined as
$\vec j_\mu^{\cal N}=
g_m\sum\limits_{k=1}^{\cal N}\vec q_{i_k}
\oint dz_\mu^k(\tau)\delta\left(x-x^k(\tau)\right)$. The couplings $g$ and $g_m$ are now dimensionless and obey the same condition $gg_m=4\pi$
as in the 3D-case. We have also parametrized the trajectory of the
$k$-th monopole by the vector $x_\mu^k(\tau)=y_\mu^k+z_\mu^k(\tau)$,
where $y_\mu^k=\int\limits_{0}^{1}d\tau x_\mu^k(\tau)$ is the
position of the trajectory, whereas the vector $z_\mu^k(\tau)$
describes its shape, both of which should be averaged over.
The fugacity of a single-monopole loop, $\zeta$, entering eq.~(\ref{Z4D}), has the dimensionality $[{\rm mass}]^4$,
$\zeta\propto {\rm e}^{-S_{\rm mon}}$, where
the action of a single $k$-th loop, obeying the
estimate $S_{\rm mon}\propto\frac{1}{g^2}\int\limits_{0}^{1}d\tau
\sqrt{\left(\dot z^k\right)^2}$, is assumed to be of the same order of magnitude
for all loops. Finally, in eq.~(\ref{Z4D}), $D_0(x)=1/(4\pi^2x^2)$ is the 4D Coulomb propagator, and the average over monopole loops
is defined similarly to eq.~(\ref{monav}) as follows:

$$\left<{\cal O}\right>_{\rm mon}=
\prod\limits_{n=0}^{\cal N}\int d^4y_n
\sum\limits_{i_n=\pm 1,\ldots,\pm\frac{N(N-1)}{2}}^{}\left<{\cal O}\right>_{z_n(\tau)}.$$
The particular form of the path-integral average over the shapes of the loops, $z_n(\tau)$'s, here is immaterial for the final (UV-cutoff dependent)
expression for the partition function (see e.g. ref.~\cite{jh} for a similar situation in the plasma of closed dual strings in the Abelian-Higgs--type
models). The only thing which matters is the normalization $\left<1\right>_{z_n(\tau)}=1$, that will be implied
henceforth. The analogue of the partition function~(\ref{II})-(\ref{III}) then reads

$$
{\cal Z}_{\rm mon}\left[\vec A_\mu\right]=\int {\cal D}\vec j_\mu\exp\left\{-{\cal S}
\left[\vec j_\mu, \vec A_\mu\right]\right\}\times$$

\begin{equation}
\label{interm}
\times\int{\cal D}\vec\chi_\mu
\exp\left\{\int d^4x\left[2\zeta\sum\limits_{i=1}^{N(N-1)/2}
\cos\left(\frac{\vec q_i\left|\vec\chi_\mu\right|}{\Lambda}\right)+i\vec\chi_\mu\vec j_\mu\right]\right\},
\end{equation}
with the action ${\cal S}$ given by eq.~(\ref{Sini}),
while the full partition function is

$${\cal Z}=\int {\cal D}\vec A_\mu{\rm e}^{-\frac14\int d^4x\vec F_{\mu\nu}^2}{\cal Z}_{\rm mon}\left[\vec A_\mu\right].$$
In eq.~(\ref{interm}),

\begin{equation}
\label{absv}
\left|\vec\chi_\mu\right|\equiv
\left(\sqrt{\chi_\mu^1\chi_\mu^1},\ldots,\sqrt{\chi_\mu^{N-1}\chi_\mu^{N-1}}\right),
\end{equation}
and $\Lambda\sim\left|y^k\right|/\left(z^k\right)^2
\gg \left|z^k\right|^{-1}$ is the UV cutoff.
Clearly, unlike the 3D-case without the $\theta$-term, the $\vec A_\mu$-field is now coupled to $\vec j_\mu$, making
${\cal Z}_{\rm mon}$ $\vec A_\mu$-dependent. This eventually leads to the change of the mass of the dual photon $\vec\chi_\mu$, as well
as to the appearance of the string $\theta$-term.

Let us address the fundamental case first.
After the saddle-point integration over $\vec\chi_\mu$,
the analogue of eqs.~(\ref{WwW})-(\ref{Vh}) takes the form

\begin{equation}
\label{danuovo}
W_a=
\frac{1}{{\cal Z}_{\rm mon}}
\int{\cal D}\vec h_{\mu\nu}
\exp\Biggl\{-S\left[\vec h_{\mu\nu}\right]
+\frac{ig}{2}\vec\mu_a\int\limits_{\Sigma(C)}^{}d\sigma_{\mu\nu}\vec h_{\mu\nu}\Biggr\},
\end{equation}
where the constraint $\partial_\mu\vec h_{\mu\nu}=0$ was already abolished. Here, the Kalb-Ramond action reads

\begin{equation}
\label{fourkalb}
S\left[\vec h_{\mu\nu}\right]=\int d^4x\Biggl(\frac{1}{4}\vec h_{\mu\nu}^2-\frac{i\theta g^2}{32\pi^2}
\vec h_{\mu\nu}\tilde{\vec h}_{\mu\nu}\Biggr)+V\left[\vec h_{\mu\nu}\right].
\end{equation}
The potential $V$ here is given by eq.~(\ref{Vh}) with the symbol $\int d^3x$ replaced by $\int d^4x$, and

$$H_a=\frac{g\Lambda}{\zeta(N-1)}\vec\mu_a\left|\partial_\mu\tilde{\vec h}_{\mu\nu}\right|.$$
In these formulae,
$\tilde{\cal O}_{\mu\nu}\equiv\frac12\varepsilon_{\mu\nu\lambda\rho}
{\cal O}_{\lambda\rho}$,
and the absolute value is defined in the same way as in eq.~(\ref{absv}), i.e., again with respect to the Lorentz indices only.
Note that the form, which the $\theta$-term has been transformed to, is quite natural, since
the respective initial expression of eq.~(\ref{Sini}) can be rewritten modulo full derivatives as

$$
\frac{i\theta g^2}{8\pi^2}
\int d^4x\vec A_\mu\vec j_\mu^{\cal N}=-\frac{i\theta g^2}{32\pi^2}\int d^4x
\left(\vec F_{\mu\nu}+\vec F_{\mu\nu}^{\cal N}\right)\left(
\tilde{\vec F}_{\mu\nu}+\tilde{\vec F}_{\mu\nu}^{\cal N}\right),$$
where [cf. eq.~(\ref{F})]

$$
\vec F_{\mu\nu}^{\cal N}(x)=-\varepsilon_{\mu\nu\lambda\rho}\partial_\lambda\int d^4yD_0(x-y)\vec j_\rho^{\cal N}(y),~~
\partial_\mu\tilde{\vec F}_{\mu\nu}^{\cal N}=\vec j_\nu^{\cal N}.$$
Next, the mass of the Kalb-Ramond field, equal to
the Debye mass of the dual photon,
which follows from the action~(\ref{fourkalb}), reads $m_D=\frac{g\eta}{4\pi}\sqrt{\left(\frac{4\pi}{g^2}\right)^2+\left(\frac{\theta}{2\pi}\right)^2}$,
where $\eta\equiv\sqrt{N\zeta}/\Lambda$. In the extreme strong-coupling limit, $g\to\infty$, this expression demonstrates an important
difference of the case $\theta=0$ from the case $\theta\ne 0$. Namely, since $\eta(g)\propto {\rm e}^{-{\rm const}/g^2}\to 1$,
$m_D\to 0$ at $\theta=0$, whereas $m_D\to\infty$ at $\theta\ne 0$. In another words, in the extreme strong-coupling limit, the correlation
length of the vacuum, equal to $1/m_D$, goes large (small) at $\theta=0$ ($\theta\ne 0$).

In the weak-field limit, $\left|H_a\right|\ll 1$, eq.~(\ref{danuovo}) yields [cf. eq.~(\ref{Wwf})]:

$$
W\left(C,\Sigma\right)_{\rm weak-field}^{\rm tot}=$$

$$=\frac{1}{{\cal Z}^{\rm tot}}
\int {\cal D}\vec h_{\mu\nu}
\exp\Biggl\{-\int d^4x\Biggl[\frac{g^2}{12\eta^2}\vec H_{\mu\nu\lambda}^2+
\frac{1}{4}\vec h_{\mu\nu}^2-\frac{i\theta g^2}{32\pi^2}
\vec h_{\mu\nu}\tilde{\vec h}_{\mu\nu}-\frac{ig}{2}\vec\mu_a
\vec h_{\mu\nu}
\Sigma_{\mu\nu}\Biggr]\Biggr\},$$
where again ${\cal Z}^{\rm tot}$ is the same integral over $\vec h_{\mu\nu}$, but with $\Sigma_{\mu\nu}$ set to zero.
Integrating over $\vec h_{\mu\nu}$, we then obtain:

$$W\left(C,\Sigma\right)_{\rm weak-field}^{\rm tot}
=\exp\Biggl\{-\frac{N-1}{4N}\Biggl[g^2\oint\limits_{C}^{}dx_\mu
\oint\limits_{C}^{}dx'_\mu D_{m_D}(x-x')+$$

\begin{equation}
\label{w4d}
+\frac{\eta^2}{2}\int d^4xd^4x'
D_{m_D}(x-x')\left(\Sigma_{\mu\nu}(x)\Sigma_{\mu\nu}(x')+
\frac{i\theta g^2}{8\pi^2}\Sigma_{\mu\nu}(x)\tilde\Sigma_{\mu\nu}(x')
\right)\Biggr]\Biggr\},
\end{equation}
where $D_{m_D}(x)\equiv m_DK_1\left(m_D|x|\right)/(4\pi^2|x|)$ is the 4D Yukawa propagator with $K_1$ denoting the
modified Bessel function. [Note that at $\theta=0$, eq.~(\ref{w4d}) takes the form of eq.~(\ref{WCSig}).]
Further curvature expansion of the
$\Sigma_{\mu\nu}\times\tilde\Sigma_{\mu\nu}$ interaction
[analogous to the expansion of the $\Sigma_{\mu\nu}\times\Sigma_{\mu\nu}$ interaction in the action~(\ref{snonloc})] yields the
string $\theta$-term equal to $ic_{\rm fund}\nu$. Here,
$\nu\equiv\frac{1}{2\pi}
\int d^2\xi\sqrt{{\sf g}}{\sf g}^{ab}(\partial_a t_{\mu\nu})(\partial_b
\tilde t_{\mu\nu})$ is the number of self-intersections of the
world sheet, and the coupling constant $c$ reads

\begin{equation}
\label{ccrit}
c_{\rm fund}=\frac{(N-1)\theta}{8N\left[
\left(\frac{4\pi}{g^2}\right)^2+\left(\frac{\theta}{2\pi}\right)^2\right]}.
\end{equation}
As it was already discussed,
$c_{\rm fund}$ is $\Lambda$-independent, since (similarly to the rigidity coupling constant $\alpha$) it is
dimensionless. We therefore see that at

\begin{equation}
\label{thcr}
\theta_{\pm}^{\rm fund}=\frac{\pi}{2}\left[\frac{N-1}{2N}\pm\sqrt{\left(\frac{N-1}{2N}\right)^2-\left(\frac{16\pi}{g^2}
\right)^2}\right],
\end{equation}
$c_{\rm fund}$ becomes equal to $\pi$, and 
self-intersections are weighted in the string partition function with the factor $(-1)^\nu$, that might
cure the problem of crumpling. This is only possible at $g\ge g^{\rm fund}\equiv 
4\sqrt{\frac{2\pi N}{N-1}}$~\footnote{Clearly, at $g=g^{\rm fund}$, $\theta_{+}^{\rm fund}=\theta_{-}^{\rm fund}=
\pi\frac{N-1}{4N}$.}, that parallels the strong-coupling regime, which should hold
in the lattice version of the model (cf. the discussion in the beginning of this Section).
In the extreme strong-coupling limit, understood in the sense $g\gg g^{\rm fund}$,
only one critical value, $\theta_{+}^{\rm fund}$, survives, $\theta_{+}^{\rm fund}\to\pi\frac{N-1}{2N}$, whereas
$\theta_{-}^{\rm fund}\to 0$, i.e., $\theta_{-}^{\rm fund}$ becomes a spurious solution, since $\left.c_{\rm fund}\right|_{\theta=0}=0$.

With the use of the results of Subsection~3.2, the adjoint-case large-$N$ counterpart of eq.~(\ref{thcr}) can readily be found.
Indeed, eq.~(\ref{ccrit}) in that case becomes replaced by

$$
c_{\rm adj}=\frac{\theta}{4\left[\left(\frac{4\pi}{g^2}\right)^2+\left(\frac{\theta}{2\pi}\right)^2\right]},
$$
so that $c_{\rm adj}$ is equal to $\pi$ at

$$
\theta_{\pm}^{\rm adj}=\frac{\pi}{2}\left[1\pm\sqrt{1-\left(\frac{16\pi}{g^2}
\right)^2}\right].$$
Note that the respective lower bound for the critical value of $g$, $g^{\rm adj}_{N\gg 1}=4\sqrt{\pi}$~\footnote{As in the 
previous footnote, at the particular value of $g$, $g=g^{\rm adj}_{N\gg 1}$, $\theta_{+}^{\rm adj}=\theta_{-}^{\rm adj}=
\frac{\pi}{2}$.}, is only slightly different 
from the value $g^{\rm fund}_{N\gg 1}=4\sqrt{2\pi}$. Similarly to the fundamental case, at $g\gg
g^{\rm adj}_{N\gg 1}$, $\theta_{-}^{\rm adj}$ becomes a spurious solution, whereas $\theta_{+}^{\rm adj}\to\pi$.
We note in conclusion that at $N\gg 1$ and in the strong-coupling limit, understood in the sense $g\gg g^{\rm fund}_{N\gg 1}$, 
the critical fundamental- and adjoint-case values of $\theta$, at which 
the problem of crumpling might be solved, are given by the following simple formula:
$\theta_{+}^{\rm fund}=\frac12\theta_{+}^{\rm adj}=\frac{\pi}{2}$.

\section{Summary}
The present paper has been devoted to the theories of confining strings, which describe the string representations
of fundamental and adjoint Wilson loops in 
the $SU(N)$-analogues of 3D and 4D compact QED. In the 3D-case, we have first found the string tension of the flat Wilson loop in the fundamental 
representation. 
This has been done by the evaluation of the mean magnetic field inside the contour of the loop, using the saddle-point analysis. 
The obtained result contains an ambiguity in the overall constant factor, that is due to the large thickness of the confining string,
which makes the definition of the mean magnetic field ambiguous. 

We have then developed, in the 3D-case, the $SU(N)$-theory of confining strings
for a Wilson loop, which belongs either to the fundamental, or to the adjoint representation. In the former case and in 
the weak-field limit of this theory (corresponding to the limit of low monopole densities), it becomes possible
to fix the above-mentioned 
factor in the string tension. Moreover, by virtue of the obtained theory one can also
find the coupling constants of the terms in the expansion 
of the nonlocal string 
effective action, which are higher in the derivatives than the Nambu-Goto term. Some of these terms, e.g. the so-called rigidity term,
vanish at the flat surface and therefore cannot be obtained upon the evaluation of the mean magnetic field inside the flat contour.
Furthermore, the theory of confining strings has been introduced 
in two various ways. One of these is to consider the Kalb-Ramond field as the field-strength tensor corresponding to the magnetic field
produced by monopoles. Free photons can further be included by abolishing a certain constraint imposed on the Kalb-Ramond field. 
In the other approach, the Kalb-Ramond field appears (with the use of the definition of the Wilson loop in the monopole plasma)
directly as the field 
unifying the monopole and free-photonic contributions to the Wilson loop.

In the case when the Wilson loop belongs to the adjoint representation, the saddle-point equation, one needs to solve for a derivation
of the confining-string theory, has a simple solution in the large-$N$ limit. In this limit, the ratio of string
couplings to the respective fundamental-case values is equal to 2. (In particular, for string tensions, this factor is the same as 
the one known in large-$N$ QCD.) 
Note also that the above-mentioned results can be readily generalized to the case of 3D Georgi-Glashow model with not-infinitely-heavy
Higgs field, along with the lines of ref.~\cite{mpla}. 

We have further considered the 4D-case with the field-theoretical
$\theta$-term included. The latter generates the string $\theta$-term which, at sufficiently strong coupling, might help in the solution
of the problem of crumpling of large world sheets. The respective (generally two, but at the particular value of the electric coupling, -- one) 
values of the $\theta$-parameter, at which this happens,
have been found in both the fundamental case and in the large-$N$ adjoint case. 
In the extreme strong-coupling limit, one of these values becomes spurious, whereas the
other one takes a very simple form.

\acknowledgments

The author is grateful for useful and encouraging discussions to Drs. M.C.~Diamantini and L.~Del~Debbio 
and to Profs. A.~Di~Giacomo, D.~Ebert,
I.I.~Kogan, F.~Lenz, Yu.M.~Makeenko, M.~Shifman, Yu.A.~Simonov, and A.~Wipf. 
He is also grateful to
the whole staff of the Physics Department of the
University of Pisa for cordial hospitality.
The work has been supported by INFN and partially by
the INTAS grant Open Call 2000, Project No. 110.

\end{document}